\documentclass{emulateapj}
\lefthead{GAUDI ET AL.} \righthead{PERIOD DISTRIBUTION OF CLOSE-IN PLANETS}
\def\eq#1{equation (\ref{#1})}

\def\days{\rm d}
\def\mj{M_{J}}
\def\rj{R_{J}}
\def\msun{M_\odot}
\def\rsun{R_\odot}
\def\ms{{\rm m\,s^{-1}}}
\def\sn{{\rm S/N}}
\def\ave#1{\left<#1\right>}
\def\cn{{\cal N}}
\def\cm{{\cal M}}
\def\cp{{\cal P}}
\def\tp{\tilde{\cal P}}
\def\ntot{N_{\rm tot}}

\begin{document}

\title{On the Period Distribution of Close-In Extrasolar Giant
Planets} \author{B.\ Scott Gaudi\altaffilmark{1}, 
S. Seager\altaffilmark{2}, and Gabriela Mallen-Ornelas\altaffilmark{1}}

\altaffiltext{1}{Harvard-Smithsonian Center for Astrophysics, 60 Garden St., Cambridge, MA 02138}
\altaffiltext{2}{Department of Terrestrial Magnetism,
Carnegie Institution of Washington, 5241 Broad Branch Rd. NW Washington, D.C. 20015}
\email{sgaudi@cfa.harvard.edu}

\begin{abstract}
Transit (TR) surveys for extrasolar planets have recently uncovered a
population of ``very hot Jupiters,'' planets with orbital periods of
$P\le 3~\days$.  At first sight this may seem surprising, given that
radial velocity (RV) surveys have found a dearth of such planets,
despite the fact that their sensitivity increases with decreasing $P$.
We examine the confrontation between RV and TR survey results, paying
particular attention to selection biases that favor short-period
planets in TR surveys.  We demonstrate that, when such biases and
small-number statistics are properly taken into account, the period
distribution of planets found by RV and TR surveys are consistent at
better than the $1\sigma$ level.  This consistency holds for a large
range of reasonable assumptions. In other words, there are not enough 
planets detected to robustly conclude that the RV and TR short-period
planet results are inconsistent.  Assuming a logarithmic distribution
of periods, we find that the relative frequency of very hot Jupiters
(VHJ: $P=1-3~\days$) to hot Jupiters (HJ: $P=3-9~\days$) is $\sim 10-20\%$.
Given an absolute frequency of HJ of $\sim 1\%$, this implies that
approximately one star in $\sim 500-1000$ has a VHJ.  We also note that VHJ and
HJ appear to be distinct in terms of their upper mass limit.  We discuss the
implications of our results for planetary migration theories,
as well as present and future TR and RV surveys.
\end{abstract}

\section{Introduction}\label{sec:intro}

\begin{figure}
\epsscale{1.0} 
\plotone{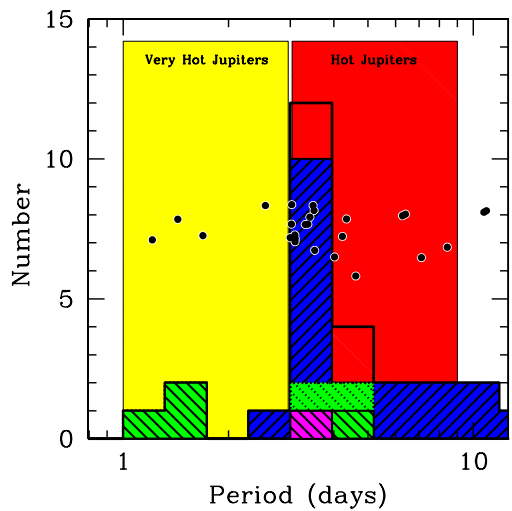}
\caption{\label{fig:one} 
The period distribution of short-period extrasolar giant planets.  
The blue shaded histogram shows planets with 
mass $M_p\sin i>0.2~\mj$ detected via radial velocity (RV)
surveys, the green shaded histogram shows planets detected
via the OGLE transit surveys, and
the magenta histogram shows the planet detected via the
TrES survey.  The dotted green histogram 
shows the periods of the two
unconfirmed candidate transiting planets with $P\ga 3~\days$ 
\citep{konacki03b}.  The unshaded histogram
shows all planets.  The yellow and red bands indicate the period
ranges for our fiducial division into very hot Jupiters and hot
Jupiters, respectively.  The black points show the individual periods
of the planets, the ordinate values are arbitrary.}
\end{figure}

Radial velocity (RV) surveys have yielded a wealth of information
about the ensemble physical properties of extrasolar planets.  This
information, in turn, provides clues to the nature of planetary
formation and evolution.  The period distribution of planets is
particularly interesting in this regard.  The very existence of
massive planets at periods of $P\la 10~\days$ was initially a
surprise.  Such planets are found around $\sim 1\%$ of main-sequence
FGK stars \citep{marcy03}, and have likely acquired their remarkable real estate via
migration through their natal disks after they accumulated the
majority of their mass.  Figure \ref{fig:one} shows the period
distribution of short-period extrasolar planets detected in RV
surveys.  We have included companions with $ M_p\sin i> 0.2~\mj$ and
$P\la 10~\days$, corresponding to velocity semi-amplitudes of
$K\ga 20~{\rm m~s^{-1}}$ for solar-mass primaries and circular orbits; 
we expect RV surveys in this region of
parameter space to be essentially complete.  Significantly, roughly
half of the 19 planets in this sample with $P\la 10~\days$ have periods in the
range $P\simeq 3-3.5~\days$.  There is a sharp cutoff below this
pile-up of planets, and there is only one planet with $P< 3~\days$,
the companion to HD73256 with $P\simeq 2.5~\days$.\footnote{Here and
throughout, we will assume for simplicity that the companion to
HD83443, which has a best-fit period of $P=2.98565\pm 0.00003$
\citep{mayor04}, actually has a period of $P=3~\days$.}  This planet
is $\sim 3$ standard deviations away from the clump of planets in the
range of $P=3-3.5~\days$, and so may be distinct in terms of its
genealogy.  Because RV surveys are likely to be substantially
incomplete for planets with mass $M_p\sin i\la 0.1~\mj$, we do not consider
the recent RV discoveries of Neptune-mass planets with periods of
$P=2.644~\days$ (GJ 436b; \citealt{butler04}), $P=2.808~\days$ (55
Cnc e; \citealt{mcarthur04}), and $P=9.55~\days$ ($\mu$ Arae c;
\citealt{santos04}).

RV surveys have so far been the most successful extrasolar planet
detection technique. Recently, two other planet detection techniques
have finally come to fruition, namely transit (TR) and microlensing surveys
\citep{bond04}.  In particular, RV follow-up of low-amplitude
transits detected by the OGLE collaboration
\citep{udalski02a,udalski02b,udalski02c,udalski03} has yielded four
bona-fide planet detections
\citep{konacki03a,bouchy04,konacki04,pont04}, and several strong
candidates \citep{konacki03b}. Recently, the Trans-Atlantic Exoplanet
Survey (TrES) collaboration announced the detection of a transiting
planet around a relatively bright K0V star \citep{alonso04}. Figure
\ref{fig:one} shows the period distribution of both the confirmed and
candidate TR-detected planets, and Table \ref{tab:data} summarizes
their properties.  Notably, the first three planets detected via
transits all have $P\simeq 1~\days$, considerably smaller than the
periods of any planets detected via RV, and well below the pile-up and
abrupt cutoff seen in the RV period distribution (see Figure
\ref{fig:one}).  This is perhaps surprising because the sensitivity of
RV surveys increases with decreasing period.

This apparent tension between the results of TR and RV surveys begs
the question of whether the results from the two techniques are
mutually consistent.  In this paper we answer this question by
considering a simple model for both the statistics and selection
biases of the TR and RV surveys.

\begin{figure}
\epsscale{1.0} 
\plotone{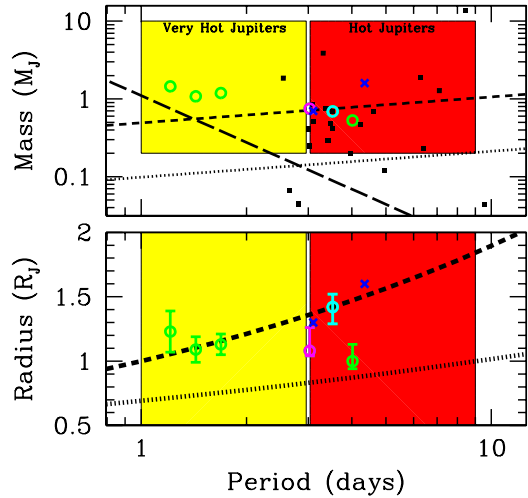}
\caption{\label{fig:two}
Physical properties of short-period extrasolar planets.  
See Table \ref{tab:data} and references therein.  Top Panel:
The points show the mass $M_p$ (or $M_p\sin i$) of short-period planets in Jupiter
masses ($\mj$) versus their period $P$ in days. Solid black points
are RV-detected planets, green circles are confirmed TR-detected
planets, blue crosses are candidate TR-detected planets, the cyan
circle is the RV-detected transiting planet HD209458b,
and the magenta circle is the bright transiting
planet TrES-1.  The dotted line
line shows $M_p\sin i$ versus $P$ for an RV semi-amplitude of $K=20~{\ms}$ for
a planet in a circular orbit and a primary of $M=\msun$.  We
assume that RV surveys are complete above this limit, and therefore
are complete to masses $M_p\sin i\ge 0.2\mj$ for $P\le 9~\days$.  
The yellow and red bands indicate the period and mass ranges for our
division into complete samples of very hot Jupiters and hot Jupiters,
respectively. The short-dashed line corresponds to $K=100~{\ms}$,
roughly appropriate for the follow-up of fainter TR candidate planets.  The
long-dashed line shows the Roche limit for a planet with radius
$R_p=2\rj$.  Bottom Panel: The points show $R_p$ versus $P$ for known
transiting planets and candidates.  Symbol types are as in the top
panel.  The dashed line shows the contour of equal number of target
stars in the effective TR survey volume, normalized to the number at $P=1~\days$
and $R_p=\rj$.  The dotted line shows the lower limit on $R_p$ versus
$P$ required to achieve a total signal-to-noise of $\sn=9$, for typical
light curves from the OGLE surveys.  See text for details.  }
\end{figure}

\begin{deluxetable*}{c|cccc|cccc|c|c}
\tablecaption{\sc  Parameters of Confirmed and Candidate Transiting Planets}
\tablewidth{0pt}
\tabletypesize{\scriptsize}
\tablehead{
  \colhead{Name} &
  \colhead{$P$ (days)} &
  \colhead{$a$ (AU)} &
  \colhead{$M_p$ ($\mj$)} &
  \colhead{$R_p$ ($\rj$)} &
  \colhead{$M_*$ ($M_\odot$)} &
  \colhead{$R_*$ ($R_\odot$)} &
  \colhead{$I$} &
  \colhead{$(V-I)$} &
  \colhead{$N_{tr}$} &
  \colhead{Reference} 
}
\startdata
OGLE-TR-56  & 1.2119 & 0.023 & $1.45\pm 0.23$ & $1.23 \pm 0.16$ & $1.04 \pm 0.05$ & $1.10 \pm 0.10$ & 15.30 & 1.26 & 11 & 1,2\\
OGLE-TR-113 & 1.4325 & 0.023 & $1.08\pm 0.28$  & $1.09\pm 0.10$   & $0.79 \pm 0.06$ & $0.78 \pm 0.06$ & 14.42 & --  & 10 & 3,4\\
OGLE-TR-132 & 1.6897 & 0.031 & $1.19\pm 0.13$  & $1.13\pm 0.08$ & $1.35\pm0.06$ & $1.43\pm 0.10$ & 15.72  &  -- &  11 & 3,5\\
OGLE-TR-111 & 4.0161 & 0.047  & $0.53\pm 0.11$  & $1.00_{-0.06}^{+0.13}$ & $0.82_{-0.02}^{+0.15}$ & $0.85_{-0.03}^{+0.10}$ & 15.55 & -- & 9 & 6 \\
\hline
OGLE-TR-10\tablenotemark{a} & 3.1014 &   -- &  $0.70\pm 0.30$ & 1.3  &   --  &    --  &  14.93	& 0.85 &  4 & 7 \\
OGLE-TR-58\tablenotemark{a} & 4.34 &   -- &  $1.60\pm 0.80$ & 1.6  &   --  &    --  &  14.75  & 1.20 &  2 & 7 \\
\hline
HD209458   & 3.5248 & 0.045  & $0.69\pm 0.05$ & $1.42_{-0.13}^{+0.10}$ & $1.06\pm 0.13$ & $1.18\pm 0.10$ & -- & -- & -- & 8,9 \\  
TrES-1     & 3.0301 & 0.039  & $0.75\pm 0.05$ & $1.08_{-0.04}^{+0.18}$ & $0.88 \pm 0.07$ & $0.85_{-0.05}^{+0.10}$ & 10.64\tablenotemark{b} & 1.15\tablenotemark{b} & -- & 10 \\
\enddata
\tablenotetext{a}{Candidate (unconfirmed) planets.}
\tablenotetext{b}{Estimated from the observed $V$-magnitude and
$J-H$ color.}
\tablerefs{
(1) \citealt{konacki03a}; (2)\citealt{torres04}; (3)\citealt{bouchy04}; 
(4)\citealt{konacki04}; (5) \citealt{moutou04}; (6)\citealt{pont04}; (7)\citealt{konacki03b}; (8)\citealt{brown01}; (9)\citealt{cs02}; (10)\citealt{alonso04} }\label{tab:data}
\end{deluxetable*}

\section{A Simple Argument}\label{sec:simple}

In this section, we present a simple, straightforward argument for why
we conclude that RV and TR surveys are essentially consistent.  These
arguments are presented in more detail in \S\ref{sec:bias}, 
\S\ref{sec:rvvtr}, and the Appendix.

The primary difference between RV and transit surveys is in how their
target stars are chosen.  RV surveys are essentially `volume-limited,'
and thus have a fixed number of target stars in their sample.  Because
RV surveys have a fixed sample size, their relative sensitivity as a
function of the mass and period depends only on the intrinsic
sensitivity of the RV technique.  This scales as $K\propto M_p\sin i
P^{-1/3}$, where the semi-amplitude $K$ characterizes the signal
strength.  It is possible to define a complete sample of planets by
considering an appropriate limit on $K$.  RV surveys are expected to
be essentially complete for $K\ga 20~{\rm m~s^{-1}}$ \citep{tt02}, which
corresponds to $M_p\sin i> 0.2~\mj$ and $P\la 9~\days$
for solar-mass primaries and circular orbits.  RV surveys indicate
that the relative frequency of $1-3~\days$ planets to $3-9~\days$
planets in this complete sample is $\sim 0.07_{-0.04}^{+0.09}$, where
the errors account for Poisson fluctuations (and are
calculated in \S \ref{sec:rvvtr}).

Field TR surveys, in contrast to RV surveys, are signal-to-noise
($\sn$) limited.  As a result, the effective volume probed by TR
surveys, and therefore the number of target stars, depends on the
total signal-to-noise ratio of the transits, which in turn depends
on the radius and period of the planets.  The basic scaling of the
sensitivity of TR surveys with period can be understood as
follows.  The flux $F$ of a star is $F \propto d^{-2}$, where $d$ is
the distance to the star.  The photometric error $\sigma \propto F^{-1/2} \propto
d$.  The number of data points $N_p$ during transits is proportional
to the duty cycle, which is inversely proportional to the semi-major
axis $a$.  Thus $N_p \propto a^{-1} \propto P^{-2/3}$.  The total
signal-to-noise of a transiting planet is $\sn \propto N_p^{1/2}
\sigma^{-1}$ and thus, at fixed $\sn$, $\sigma \propto N_p^{1/2}
\propto P^{-1/3}$.  Therefore, $d\propto P^{-1/3}$, i.e.\ the distance
out to which one can detect a transiting planet at fixed
signal-to-noise scales as $P^{-1/3}$.  The number of stars in the
survey volume is $\propto d^{3} \propto P^{-1}$.  Combined with the
transit probability, which scales as $a^{-1} \propto P^{-2/3}$, this
implies an overall sensitivity $\propto P^{-5/3}$. 

Thus, TR surveys are,
on average,  $\sim (1/3)^{-5/3} \sim 6$ times more sensitive
to $P=1~\days$ planets than $P=3~\days$ planets.  The observed
relative frequency of confirmed $1-3~\days$ to $3-9~\days$ planets
discovered in the OGLE TR surveys is $\sim 3$, which corresponds to an
intrinsic relative frequency (after accounting for the factor of $6$)
of $\sim 0.5_{-0.3}^{+1.5}$, as compared to  $\sim 0.07_{-0.04}^{+0.09}$,
for the RV surveys.  Thus, considering the
large errors due to small number statistics, RV and TR surveys are
basically consistent (at better than the $\sim 2\sigma$ level).  If at least one
of the remaining OGLE $P\ge 3~\days$ planet candidates is confirmed
in the future, then TR
and RV surveys are consistent at better than $1\sigma$.  In other
words, there are not enough planets detected to robustly conclude
that the RV and TR short period planet results are inconsistent.

As we discuss in more detail in the the Appendix, there are additional
effects that favor the {\it confirmation} of shorter-period transiting
planets.  First, shorter-period planets will generally tend to exhibit
more transits; this makes their period determinations from the TR
data more accurate.  Accurate periods aid significantly in RV
follow-up and confirmation.  Second, shorter-period planets will
generally have larger velocity semi-amplitudes $K$, both because of
their smaller periods $(K\propto P^{-1/3})$, and because there appears
to be a dearth of massive ($M_p\sin i\ga \mj$) planets with $P=3-9~\days$ (see
Figure \ref{fig:two}).

\section{Selection Effects in Transit Surveys}\label{sec:bias}

In this section, we present a more detailed derivation
of the sensitivity of signal-to-noise limited
planet TR surveys as a function of the period and radius
of the planet.  

Field searches for transiting planets are very different from RV
searches as they are uniform surveys, in which the target stars are all
observed in the same manner (rather than targeted observations of
individual stars).  Therefore, the noise properties vary from star to
star.  As a result, the relative number of planets above a given $\sn$
threshold depends not only on the way in which the intrinsic signal
scales with planet properties, but also on the number of stars with a
given noise level.  Since, for transit surveys, the noise depends on
the flux of the star, which depends on the distance to the star, the
effective number of target stars depends on the number of stars in the
effective survey volume that is defined by the maximum distance out to
which a planet produces a $\sn$ greater than the threshold.  This
leads to a strong sensitivity of TR surveys on planet period and
radius (as well as parent star mass and luminosity, see
\citealt{pgd03}), which we now derive.

The total signal-to-noise of a transiting planet can be approximated as
\begin{equation}
{\rm \frac{S}{N}}= N_{p}^{1/2}
\left(\frac{\delta}{\sigma}\right).  
\label{eqn:sntrans}
\end{equation}
Here $N_p$ is the total number of measurements during the transit,
$\delta$ is the depth of the transit, and $\sigma$ is the fractional
flux error for a single measurement.  We can approximate $N_p = \ntot(R_*/\pi a)$ (for a central transit)
and $\delta = (R_p/R_*)^2$.  Here $\ntot$ is
the total number of observations, $a$ is the semi-major axis of the
planet, and $R_*$ is the radius of the parent star.   Combining
these relations with Kepler's third law, we have
\begin{equation}
{\rm \frac{S}{N}} \propto P^{-1/3} R_p^2 \sigma^{-1}
\label{eqn:snprops}
\end{equation}

We then estimate the relative sensitivity as follows.  Following 
\citet{pgd03},
the number of target stars for which a planet of a given $P$ and $R_p$ would
produce a $\sn$ greater than a given threshold is proportional to
\begin{equation}
\frac{{\rm d}^2\cn(P,R_p)}{{\rm d}P {\rm d}R_p} \propto f(P,R_p)
 P_T(P) V_{\rm max}(P,R_p),
\label{eqn:ndet}
\end{equation}
where $f(P,R_p)\equiv {{\rm d}^2 n(P,R_p)}/{{\rm d}P {\rm d}R_p}$ is the intrinsic frequency of planets
as a function of $P$ and $R_p$, $P_T(P)$ is the probability that a planet of a given $P$ will
transit its parent star, and $V_{\rm max}(P,R_p)$ is the maximum
volume within which a planet of a given $P$ and $R_p$ can be detected.
The geometric transit probability is simply $P_T \simeq R_*/a \propto P^{-2/3}$.  
We assume the form $V_{\rm max} \propto F_{\rm min}^{-3/2}$, where $F_{\rm
min}(P,R_p)$ is the minimum flux of a star around which 
a planet of period $P$ and radius 
$R_p$ can be detected; this form is
appropriate for a constant volume density of stars and no
extinction.
For fixed $\sn$, we have from
\eq{eqn:snprops} that $\sigma \propto P^{-1/3} R_p^2$.  For
source-dominated photon noise, $\sigma
\propto F_{\rm min}^{-1/2}$, and so $F_{\rm min} \propto P^{2/3} R_p^{-4}$
and $V_{\rm max} \propto P^{-1} R_p^{6}$.  
Finally, combining this with $P_T \propto P^{-2/3}$, we find
\begin{equation}
\frac{{\rm d}^2\cn}{{\rm d}P {\rm d}R_p} \propto  f(P,R_p) R_p^{6}P^{-5/3}.
\label{eqn:ndetrpa}
\end{equation}
This strong function of $P$ implies that the TR surveys are
very biased toward detecting short-period planets.

Note that, in deriving \eq{eqn:ndetrpa}, we have made the simplistic
assumption that the number of data points during transit is
proportional to the duty cycle, $N_p \propto R_*/\pi a$.  This
assumes random sampling and short periods as compared to the transit
campaign.  In fact, actual transit campaigns have non-uniform sampling
and finite durations.  In addition, transit candidates require RV
follow-up for confirmation; this introduces additional selection
effects.  We consider both effects in detail in the Appendix.

\section{Radial Velocity Versus Transits}\label{sec:rvvtr}

We now address the question of whether the period distribution of the
planets discovered by RV and TR surveys are consistent, considering both the
selection biases discussed in the previous section, as well as the
effects of small-number statistics.

For our fiducial comparisons, we consider two equal-width logarithmic
bins in period with $(P_{1,\rm min},P_{1,\rm max})=(1\days,3\days)$
and $(P_{1,\rm min},P_{1,\rm max})=(3\days,9\days)$.  We argued in
\S\ref{sec:bias} that local RV surveys should be essentially complete
for planets with velocity semi-amplitude $K \ga 20~\ms$, and therefore
if we restrict our analysis to $M_p\sin i\ge 0.2\mj$, then the observed
number of planets detected by RV in these two bins should be an
unbiased sample of the true distribution of planets (see Figure
\ref{fig:two}).  We will also restrict our attention to $M_p\sin i \le
10\mj$ to avoid possible brown dwarf candidates.  The number of RV
planets with $0.2\mj \le M_p\sin i \le 10\mj$ in our two fiducial period
bins is given in Table 2.  There is one planet in the first bin, and
15 in the second.  Therefore the relative frequencies are
$\sim 7\%$.  We denote these two complete samples as ``Very Hot
Jupiters'' (VHJ) and ``Hot Jupiters'' (HJ), respectively.

For comparison to the RV surveys, we will consider the results from
the two campaigns by the OGLE collaboration.\footnote{In this section,
we will not
consider the recently-detected bright transiting planet TrES-1
\citep{alonso04}, since the survey details necessary
for the statistical analysis are not available.}
%this survey has as yet yielded only one planet.}  
Pertinent details about the OGLE surveys are summarized in
the Appendix.  Because the OGLE searches are signal-to-noise limited
surveys, as opposed to the volume-limited RV surveys, it is not
possible to define a complete, unbiased sample of observed planets
(see the discussion in \S\ref{sec:bias}), and we must take into
account the selection biases to infer the true planet frequency.  We first
assume a frequency distribution $f(P,R_p)$.  We assume that planets
are uniformly distributed in $\log{P}$ within each bin, and that all
planets in bin $i$ have radius $R_{p,i}$.  This gives an intrinsic
frequency distribution of,
\begin{equation}
f_i(P,R_p)=\frac{{\rm d}^2 n_i(P,R_p)}{{\rm d}P {\rm d}R_p}
= \frac{N_i}{\Delta {\log P_i}} 
P^{-1} \delta_{\rm D}(R_p-R_{p,i})
\label{eqn:intrfreq}
\end{equation}
where $N_i$ is the total number of planets in bin $i$, 
$\Delta {\log P_i}=\log P_{i,\rm max}-\log P_{i,\rm min}$ is logarithmic width
of the bin, and $\delta_{\rm D}$ is the Dirac delta function.  
From equation \ref{eqn:ndetrpa}, the expected number $\cn_i$ of observed
transiting planets in bin $i$ is,
\begin{equation}
\cn_i(P,R_p) \propto \int {\rm d}R_p \int_{P_{i,\rm min}}^{P_{i,\rm max}} 
{\rm d}P\, f_i(P,R_p) R_p^{6}P^{-5/3}.
\label{eqn:cni}
\end{equation}
The constant of proportionality is independent of $P$ and $R_p$,
and thus the ratio of the observed number of planets in the two bins is simply, 
\begin{equation}
\frac{\cn_1}{\cn_2} = 
r_{12}
\left( \frac{R_{p,1}}{R_{p,2}} \right)^6 
\left( \frac{P_{1,\rm min}}{P_{2,\rm min}} \right)^{-5/3},
\label{eqn:cn1cn2}
\end{equation}
where we have assumed that $\Delta \log P_1=\Delta \log P_2$, 
and we have defined $r_{12}=N_1/N_2$, the ratio of the 
intrinsic number of planets in the two period bins, i.e.\ the relative
frequency of VHJ and HJ.

For simplicity, we will assume that VHJ and HJ have
similar radii on average, and so $R_{p,1}=R_{p,2}$.  For our
period bins, the last factor is $(1\days/3\days)^{-5/3}\sim 6$.  The
number of planets detected by TR surveys in the first bin is $\cn_1
=3$.  There is one confirmed OGLE planet detected by TR in the second bin.
This implies a intrinsic relative frequency of VHJ and HJ
of $r_{12} \sim 50\%$, which is a factor of $\sim 7$ larger than 
inferred from RV surveys.    

Given the relatively small number of planets in each of our two fiducial
period bins, we must account for Poisson fluctuations in order to
provide a robust estimate of the relative frequency $r_{12}$.  
In the limit of a large number of trials, the probability $\cp$
of observing $\cn$ planets given $\cm$ expected planets is,
\begin{equation}
\cp(\cn|\cm)=\frac{ e^{-\cm}\cm^\cn}{\cn!}.
\label{eqn:poisson}
\end{equation}
For large $\cm$, the probability of observing any particular value of $\cn$
becomes small, simply because of the large number of possible outcomes.  We 
therefore consider relative probabilities 
$\tp(\cn|\cm)\equiv \cp(\cn/\cm)/\cp_{\rm max}(\cm)$, and normalize
$\cp(\cn|\cm)$
by the maximum probability $\cp_{\rm max}(\cm)\equiv{\rm max}[\cp(1|\cm),\cp(2|\cm),...,\cp(\infty|\cm)]$ for a given expected number $\cm$.\footnote{Rather than considering 
relative probabilities, one might instead consider cumulative probabilities $\cp(<\cn|\cm)$. 
We find that these two approaches yield similar results.}

We can now construct probability distributions $\cp(r_{12}|\cn_1,\cn_2)$ of $r_{12}$, 
given the observed numbers $\cn_1$ and
$\cn_2$ of VHJ and HJ, and incorporating selection biases and Poisson
fluctuations.  This probability is,
\begin{equation}
\cp(r_{12}|\cn_1,\cn_2) \propto \int {\rm d}\cm_1 \tp(\cn_1|\cm_1) \tp(\cn_2|\cm_2),
\label{eqn:pr12}
\end{equation}
here $\cm_2$ depends on $\cm_1$ and $r_{12}$.  For RV, it is
simply $\cm_2=\cm_1/r_{12}$, whereas for TR, it is related via
\eq{eqn:cn1cn2} (replacing $\cn \rightarrow \cm$).  Note that, up to a
constant, \eq{eqn:pr12} is equivalent under the transposition $\cm_1 \leftrightarrow \cm_2$,
and we could have also integrated over $\cm_2$.

\begin{figure}
\epsscale{1.0} 
\plotone{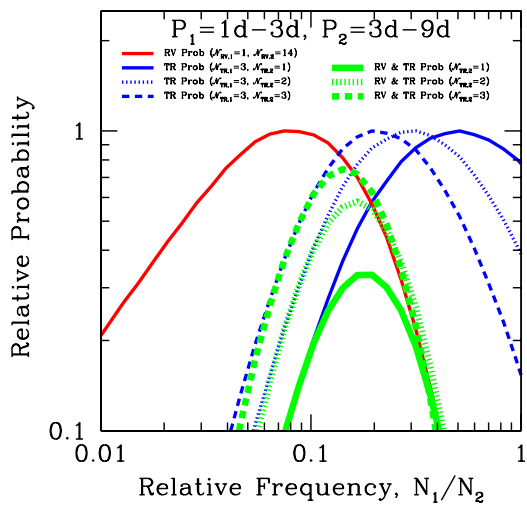}
\caption{\label{fig:three} 
The lines show the relative probability of observing $\cn_1$ planets
in the period range $P_1=$1d-3d and $\cn_2$ planets in the period
range $P_1=$3d-9d, for an absolute relative frequency of planets in
these two period ranges of $r_{12}\equiv N_1/N_2$, and assuming a uniform
logarithmic distribution in $P$ in each bin.  The curves take into
account both Poisson fluctuations and period-dependent selection
biases.  The red curve shows the probability corresponding to the RV surveys,
which observe $\cn_{{\rm RV},1}=1$ in the period range $P_1$=1d-3d and
$\cn_{{\rm RV},2}=15$ in the period range $P_2$=3d-9d.  The blue
curves are for the TR surveys, for $\cn_{{\rm TR},1}=3$, and
$\cn_{{\rm TR},2}=1$ (solid), 2 (dotted),
and 3 (dashed).  The green
curves are joint RV and TR probabilities, for $\cn_{{\rm RV},1}=1$,
$\cn_{{\rm RV},2}=14$, $\cn_{{\rm TR},1}=3$, and $\cn_{{\rm TR},2}=1$
(solid), 2 (dashed), or 3 (dotted).
}
\end{figure}

\begin{deluxetable*}{c|ccc|ccc|cc}
\tablecaption{\sc  Radial Velocity versus Transits}
\tablewidth{0pt}
\tabletypesize{\scriptsize}
\tablehead{
  \colhead{Assumption\tablenotemark{a}} &
  \colhead{Range} &
  \colhead{\# RV} &
  \colhead{\# TR} &
  \colhead{Range} &
  \colhead{\# RV} &
  \colhead{\# TR} &
  \colhead{$N_{1}/N_{2}$\tablenotemark{b}} &
  \colhead{Prob.\tablenotemark{c}}
}
\startdata
Logarithmic & $P$=1d-3d & 1  & 3  & $P$=3d-9d & 15 & 0 & $0.21_{-0.09}^{+0.16}$ & $12.7\%$ \\
 	 -- & -- 	& -- & -- & -- 	       & -- & 1 & $0.18_{-0.08}^{+0.12}$ & $33.1\%$ \\
	 -- & -- 	& -- & -- & -- 	       & -- & 2 & $0.15_{-0.07}^{+0.10}$ & $54.1\%$ \\
	 -- & -- 	& -- & -- & -- 	       & -- & 3 & $0.13_{-0.05}^{+0.09}$ & $70.9\%$ \\
\hline
Logarithmic & $P$=1d-2d & 0  & 3  & $P$=2d-4d  & 11 & 0 & $0.25_{-0.12}^{+0.21}$ & $2.0\%$  \\
 	 -- & -- 	& -- & -- & -- 	       & -- & 1 & $0.22_{-0.10}^{+0.18}$ & $6.5\%$ \\
	 -- & -- 	& -- & -- & -- 	       & -- & 2 & $0.20_{-0.09}^{+0.14}$ & $13.3\%$ \\
	 -- & -- 	& -- & -- & -- 	       & -- & 3 & $0.17_{-0.08}^{+0.12}$ & $20.7\%$ \\
\hline
Linear      & $P$=1d-3d & 1  & 3  & $P$=3d-5d  & 12 & 1 & $0.25_{-0.11}^{+0.18}$ & $22.6\%$ \\
\enddata
\tablenotetext{a}{Assumed form for the period distribution.}
\tablenotetext{b}{Inferred intrinsic relative frequency of VHJ and HJ, from the joint RV and TR results.}
\tablenotetext{c}{Probability of observing both RV and TR results at
the peak of the distribution of $N_1/N_2$.}
\label{tab:probs}
\end{deluxetable*}

Figure \ref{fig:three} shows the probability distribution for
$r_{12}$, normalized to the peak probability, as inferred from RV and
TR surveys, assuming that $\cn_2=1,2$ or $3$ of the candidate TR
planets with $P=3\days-9\days$ are real.  The probability
distributions peak at the expected value given the observed numbers of
VHJ and HJ.  However, due to Poisson fluctuations, the distributions
are quite broad.  For example, the RV surveys imply a median and 68\%
confidence interval of $r_{12}=0.07_{-0.04}^{+0.09}$, whereas the TR
surveys with $\cn_{{\rm TR},2}=1$ imply $r_{12}=0.5_{-0.3}^{+1.5}$.
Therefore, it is clear that when Poisson fluctuations are taken into
account, these two determinations are roughly consistent.  Figure
\ref{fig:three} also shows the product of the relative probabilities
of $r_{12}$ from the RV and TR surveys.  
Considering the one confirmed $P\ge 3\days$ OGLE planet
($\cn_{{\rm TR},2}=1$), the median and 68\% confidence interval for
the joint probability distribution is $r_{12}=0.18_{-0.08}^{+0.12}$.
 The peak probability is $\sim 33\%$.  Table 2 summarizes the
inferred values and peak probabilities of $r_{12}$, including the
cases $\cn_{{\rm TR},2}=0,1,2,3$.  Even if none of the $P>3~\days$ TR
planets were real, the TR and RV surveys would still have been
compatible at the $\sim 2\sigma$ level.  We therefore conclude that the TR
and RV surveys are consistent, and imply a relative frequency of VHJ
to HJ of $\sim 10-20\%$, with the precise number and degree
of consistency depending on how many
of the $P>3~\days$ TR planets turn out to be real.

Two of the HJ in our sample orbit stars that are members of a binary
system ($\tau$ Boo and $\upsilon$ And).  There have been various
studies that indicate that such planets may have properties that are
statistically distinct from those of planets orbiting single stars (e.g., \citealt{eggen04}).
Since it is unclear whether planets orbiting stars that are members
of a binary system could be detected in the OGLE surveys, it is
interesting to redo the analysis above, excluding these two planets.  
We find that doing so leaves our conclusions unchanged.  For example, we infer a relative
frequency of $r_{12}=0.16^{+0.12}_{-0.07}$ for $\cn_{{\rm TR},2}=2$, with a peak
probability of $62\%$, as compared to $r_{12}=0.15_{-0.07}^{+0.10}$ and a peak probability
of $54\%$ when we include these two planets.

If we include in our analysis planets with mass $M_p \sin i \le 0.2 \mj$ 
(and so the two new Neptune mass planets with
$P<3~\days$, \citep{butler04, mcarthur04}), as well as the
newly-discovered bright transiting planet TrES-1 \citep{alonso04} with
$P=3.0301~\days$, RV and TR surveys
imply relative frequencies of $0.16_{-0.08}^{+0.15}$ and
$0.32_{-0.18}^{+0.46}$, respectively.  In other words, the two 
types of surveys are
highly consistent.   Combining both surveys, we find a relative frequency of
$0.20_{-0.08}^{+0.13}$, with a peak probability of $\sim 87\%$.  We
stress that including these planets is probably not valid, because (1)
RV surveys are very incomplete for $M_p\sin i \la 0.1~\mj$, (2) it is
not at all clear that TR surveys could detect planets with mass as low
as Neptune, (3) even if the TR surveys could detect such planets, they
would be extremely difficult to confirm from follow-up RV measurements,
and (4) the details of the TrES survey necessary for a
proper statistical analysis are unknown.  However, the fact that the
relative frequency agrees with that inferred when these planets are
not included demonstrates that our conclusions are fairly robust.

We have checked that changing the binning or the form of period
distribution does not alter our conclusions substantially.  For
example, if we choose equal logarithmic bins of $1-2~\days$ and
$2-4~\days$, the RV surveys imply a $1\sigma$ upper limit to
the relative frequency of planets with $P=1-2~\days$ versus
$2-4~\days$ of $0.2$.  This is compared to a relative frequency of
$1.1_{-0.7}^{+2.8}$ implied by TR surveys.  In this case,
TR and
RV surveys are consistent at the $\sim 2\sigma$ level.  
Taken together, TR and RV surveys imply
a relative frequency of $0.22_{-0.10}^{+0.18}$
for $\cn_{{\rm TR},2}=1$, with a peak probability of $\sim 7\%$.  
For planets distributed linearly with period, and period bins of
$1-3~\days$ and $3-5~\days$, we find a relative frequency of
$0.25_{-0.11}^{+0.18}$ for $\cn_{{\rm TR},2}=1$, with a peak
probability of $\sim 23\%$.  We have also checked that
aliasing due to uneven sampling does not affect our results substantially.
See the Appendix for more details.

\section{Hidden Assumptions, Caveats and Complications}\label{sec:candc}

In this section, we briefly mention various caveats and complications
that may affect our results in detail.  We begin by making a list of
some of the more important hidden assumptions we have made.  For completeness,
we also list assumptions that we have already addressed. 

\medskip
\noindent
1. {\em $\sn$-limited TR Surveys:} We have assumed that the detection of
planets in the OGLE surveys is limited only by $\sn$, and not by,
e.g.\ apparent magnitude.  In other words, all stars for
which planets (with the periods and radii we consider) 
would produce transits with
$\sn>9$ are considered.  We discuss the validity of this assumption in
more detail below. 

\medskip
\noindent
2. {\em Uniform Sampling:} For the majority of our results, we have
assumed uniform sampling.

\medskip
\noindent
3. {\em Logarithmic Period Distribution and Specific Binning:} For the majority of our results, we have assumed a logarithmic intrinsic
period distribution, and specific choice of bins of $P=1-3~\days$ and
$P=3-9~\days$.  

\medskip
\noindent
4. {\em All Detected Planets Can Be Confirmed:} We have implicitly
assumed that all planets detected in TR surveys can be confirmed via
follow-up RV observations, regardless of their period.  Because of the
prevalence of false positives that mimic planetary transit signals, it is not possible
to use the observed relative frequency of planet {\it candidates} as a
function of period to infer the the true frequency, one must instead
use the observed frequency of true planets, as confirmed by follow-up
RV observations. 

\medskip
\noindent
5. {\em Homogeneous Stellar Populations:} We have assumed that the
population of source stars does not vary as a function of distance,
and therefore that terms in the transit sensitivity that depend on the
mass, radius, and luminosity of the host stars drop out.  

\medskip
\noindent
6. {\em Uniform Stellar Density:} We adopted $V_{\rm max} \propto
F_{\rm min}^{-3/2}$, which is assumes a constant volume density of
stars and no dust.  

\medskip
\noindent
7. {\em Uniform Intrinsic Period Distribution:} We
have assumed that the period distribution of planets is uniform
(either in log or linear period).  It is clear, given the
`pile-up' of planets at $P\sim 3~\days$, that this 
assumption cannot be correct in detail.   

\medskip
\noindent
8. {\em Photon and Source Limited Noise:}
We assumed that the photometric precision is photon-noise limited (i.e.\ 
no systematic errors), and
furthermore dominated by the source (i.e.\ sky noise is negligible).

\medskip
\noindent
9. {\em Correspondence Between Detectable RV and TR Planets:} We have
assumed that all planets in the `complete' sample from RV surveys are
detectable in TR surveys, i.e.\ that both surveys probe the same
population of planets.  

\medskip
\noindent
10. {\em Constant Radii:} We have
assumed that VHJ and HJ have equal, constant radii.  

\medskip
\noindent
11. {\em No Correlation Between Planet and Stellar Properties:} 
We have assumed that the physical properties of short-period planets are
uncorrelated with the physical properties of their parent stars.

\medskip

The first assumption, namely that the OGLE TR surveys are $\sn$-limited,
is the most crucial, as it provides the crux of
our argument that transit surveys are much more sensitive to short
period planets than long period planets.  In fact, the OGLE surveys
are not strictly $\sn$-limited, as several cuts were imposed to
preselect light curves to search for transiting planets.  
Of the cuts made, the most relevant here was the exclusion of
light curves whose root-mean-squared (RMS) scatter exceeded 1.5\%.
This is important because it effectively limits the volume which is
searched for planets, in a way that depends on the period and radius
of the planet.  If this volume is smaller than the largest volume for
which the $\sn$ is greater than the threshold, then the survey is
no longer $\sn$-limited.  From the definition of the $\sn$
(Eq.~\ref{eqn:sntrans}), and assuming that the maximum photometric error
$\sigma_{\rm max}$ is equal to the maximum RMS, 
we find that the TR surveys are
$\sn$-limited provided that the ratio of planet radius to stellar
radius satisfies,
\begin{equation}
\frac{R_p}{R_*}\le \left(\frac{\rm S}{\rm N}\right)^{1/2}
\left(\frac{\sigma_{\rm max}}{N_{\rm tot}^{1/2}}\right)^{1/2} \left(\frac{P^2 \pi GM_*}{4 R_*^3}\right)^{1/12}
\label{eqn:rplima}.
\end{equation}
For $\sigma_{\rm max}=1.5\%$ and a threshold of $\sn=9$, 
\begin{equation}
\frac{R_p}{R_*}\le 0.12
\left(\frac{P}{1\days}\right)^{1/6}
\left(\frac{M_*}{M_\odot}\right)^{1/12}
\left(\frac{R_*}{R_\odot}\right)^{-1/4}
\left(\frac{N_{\rm tot}}{10^3}\right)^{-1/4}.
\label{eqn:rplim}
\end{equation}
For the 2002 OGLE campaign, $N_{\rm tot}=1166$.  This gives for
$M_*=M_\odot$, $R_*=R_\odot$, and $P=1~\days$ (the smallest period we
consider) $R_p/R_* \la 0.12$. Therefore, for solar-type primaries, TR
surveys are not $\sn$-limited for the largest planets and smallest
periods, and the arguments we have presented that are based on this
assumption will break down.  In practice, the magnitude of the
correction will depend on the size distribution of planet radii, as
well as the distribution of primary radii.  However, if planets with
$R_p/R_* \ge 0.12$ are relatively rare, then the correction will
generally be small.  We note that the sensitivity of TR surveys to
planets around small primaries can be severely reduced by imposing
magnitude or RMS limits, and thus future transit searches should take
care when making such cuts that they are not rejecting otherwise
viable candidates.

We have discussed the effects of our second, third, and fourth
assumptions on our results in \S\ref{sec:rvvtr} and the Appendix.
Although violations of these can and do affect our results in detail,
they do not change our basic conclusions substantially.

Violations of the remaining assumptions will have various effects on
our conclusions, however investigation of these in detail is well
beyond the scope of the paper.  Furthermore, although the importance
of many of these assumptions can be determined directly from data,
these data are not presently available.  In the end, however,
our assumptions are approximately valid, and a more
careful examination of these issues is not warranted, given the small
number of detected planets and resulting poor statistics.  Our primary
goal is to provide general insight into the biases and selection
effects inherent in RV and (especially) TR surveys.  We note that,
when many more planets are detected and the present analysis
revisited, the assumptions listed above will likely have to be
reconsidered more carefully.

\section{Discussion}\label{sec:imp}

We have demonstrated that the sensitivity of signal-to-noise limited
transit surveys scales as $P^{-5/3}$.  This strong dependence on $P$
arises from geometric and signal-to-noise considerations, and implies
that transit surveys are $\sim 6$ times more sensitive to $P=1~\days$
planets than $P=3~\days$ planets.  When these selection biases and
small number statistics are properly taken into account, we find that
the populations of close-in massive planets discovered by RV and TR
surveys are consistent (at better than the $2\sigma$ level).  
In other words, there are not enough planets detected to robustly
conclude that the RV and TR short-period planet results are 
inconsistent.  
We then 
used the
observed relative frequency of planets as a function of period as
probed by both methods to show that the HJ are approximately 5-10
times more common that VHJ.  

RV surveys have demonstrated that the absolute frequency of HJ is
$\sim 1\%$ \citep{marcy03}, and thus the frequency of VHJ is 
$\sim 0.1-0.2\%$, i.e.\ 1 in 500-1000 stars have a VHJ.  The frequency of
VHJ is approximately the same as the frequency of transiting HJ, and
therefore future RV surveys that aim to detect short-period planets by
monitoring a large number of relatively nearby stars over short time
periods \citep{fischer04}  should detect VHJ at approximately the
same rate as transiting planets.  Should such RV surveys {\it not}
uncover VHJ at the expected rate, this would likely point to a
difference in the populations of planetary systems probed by RV and TR
surveys.

Roughly $15\%$ of VHJ should transit their parent stars, as opposed to
$\sim 7\%$ of HJ, and approximately one in 3300-6700 single main sequence
FGK stars should have a transiting VHJ, as opposed to one in 1400 for HJ.
It has been estimated that there are $\sim 30$ detectable transiting
HJ around stars with $V \la 10$ in the entire sky \citep{pgd03,deeg04}, and thus
$\sim 7-13$ transiting VHJ.  The detection of only $3$ VHJ in the OGLE surveys
containing $\sim 150,000$ stars implies that only $5-10\%$ of the
sources are single, main-sequence, FGK stars useful for detecting
transiting planets, roughly in accord with, but somewhat smaller than,
the fraction estimated for TR surveys of brighter stars
\citep{brown03}.  The fact that other deep surveys such as EXPLORE
\citep{explore03} have not detected any promising VHJ candidates
despite searching a similar number of stars may be due to either small
number statistics, reduced efficiency due to shorter observational
campaigns, or both.  Finally, 
we estimate that {\it Kepler} should find $\sim
15-30$ transiting VHJ around the $\sim 10^5$ main-sequence stars in its
field-of-view.

It is interesting to note that there is some evidence that VHJ and HJ
also appear to differ in their mass.  Figure \ref{fig:two} shows the
distribution of confirmed and candidate planets in the mass-period and
radius-period plane.  While there is a paucity of high-mass ($M_p\ga
\mj$) planets with periods of $P\sim 3-10~\days$ \citep{pr02,zm02},
all of the planets with $P\la 3~\days$ have $M\ge \mj$.  This includes
the RV planet HD 73256b with $P=2.5~\days$, which argues that this
planet is indeed a VHJ, and thus that RV surveys have already detected
an analog to the OGLE short-period planets.  
The lack of high-mass HJ is certainly
real, and thus the mere existence of VHJ with $M\ge \mj$ points
toward some differentiation in the upper mass limit of the two
populations.  Whether or not the lack of lower-mass $0.5\mj \la
M_p \la \mj $ VHJ is real is certainly debatable.  For TR-selected
planets, this could in principle be a selection effect if the radius
is a strong decreasing function of decreasing mass in this mass range,
however this is neither seen for the known planets with measured radii, nor
is expected theoretically.  RV follow-up would likely prove more difficult
for such lower-mass objects, however (see Figure \ref{fig:two}).

As can be seen in Figure 2, there appears to be an `edge' in
the distribution of planets in the mass-period plane that is
reasonably well-described by twice the Roche limit for a planet radius
of $R=R_J$.  This has been interpreted as evidence that
short period planets may have originated from highly-eccentric
orbits, which underwent strong tidal evolution with their parent
stars, leading to circularization at twice the Roche limit \citep{faber04}.
However, this model alone cannot explain the pile-up at 3 days and
paucity of VHJ relative to HJ.  Alternatively, it may be that massive
$M\ge M_J$ planets were not subject to whatever mechanism halted the
migration of less massive planets at periods of $P \simeq 3~\days$.
Rather, these massive planets migrated on quasi-circular orbits while
they were still young (and thus relatively large, $\sim 2\rj$), through periods of
$3~\days$, until they reached their Roche limit, at which point they may
have lost mass and angular momentum to their parent star, halting their
inward migration (e.g.\ \citealt{trilling98}).  

The recently-discovered short-period Neptune-mass planets
\citep{santos04,mcarthur04,butler04} complicate the interpretation of
the properties of short-period planets even further. Two of these
planets have periods that are less than the $3~\days$ limit observed
for planets with mass $0.2-1~\mj$.  
Both of these planets show marginal ($\sim 2\sigma$) evidence for non-zero
eccentricity.  In addition, 55 Cnc has a more distant companion with
$P=14.653~\days$ \citep{mcarthur04}, and the RV curve for GJ 436b
has marginal evidence for a linear trend, consistent with a more distant
companion.  Since tidal torques
would be expected to circularize the orbits of such close planets on
an extremely short time scale, this may be evidence for dynamical
interactions with more
distant companions, which may affect their migration and
explain why they do not obey the $3~\days$ migration limit.  However,
we stress that the evidence for non-zero eccentricity 
in these planets is marginal,
and thus will need to be confirmed with additional observations
before any firm conclusions can be drawn.  The fact
that these planets have orbits that are smaller than the Roche edge
observed for higher-mass planets is understandable if they are
primarily rocky in composition.  

It is clear that much remains to be understood about short period
extrasolar planetary companions.  In this regard, building statistics
is essential.  Future RV surveys that tailor their observations to
preferentially discover large numbers of short-period planets are very
important, and are currently being undertaken \citep{fischer04}.
Complementarity is also essential: the success of TR surveys in
uncovering a population of heretofore unknown planets demonstrates the
benefit of searching for planets with multiple methods, each of which
have their own unique set of advantages, drawbacks, and biases.  In
addition to the success of OGLE and TrES, all-sky shallow TR surveys
\citep{deeg04,pgd04,bakos04}, wide-angle field surveys
\citep{kane04,bc00}, deep ecliptic surveys \citep{explore03}, and
surveys in stellar systems \citep{mochejska02,burke03, street03, vonbraun04}
should all uncover a large number of short-period planets which can be
compared and combined with the yield of RV surveys to provide
diagnostic ensemble properties of short-period planets.

{{\it Note:} After the original submission of this paper, and during
the refereeing process, we learned of several new results that bear on
the discussion here.  \citet{bouchy05} report on their follow-up of
OGLE bulge candidates.  They argue that OGLE-TR-58, listed as a
possible planet candidate by \citet{konacki03b}, is more likely a
false positive caused by intrinsic stellar photometric variability.
They also report radial velocity measurements of OGLE-TR-10 that
indicate a possible planetary companion, in agreement with sparser RV
data from \citet{konacki03b}.  Very recently, \citet{konacki05} report
additional RV measurements of OGLE-TR-10, confirming the planetary nature of its
companion, which has a radius $R_p= 1.24 \pm 0.09 \rj$, and
a mass $M_p=0.57 \pm 0.12 \mj$.  The 
mass of this planet is consistent with other HJ, and significantly
less than that of the known VHJ, reinforcing the case for a difference in
the mass of these two populations of planets.  With the confirmation
of OGLE-TR-10, the number of HJ discovered in the OGLE transit surveys
is $\cn_{{\rm TR},2}=2$.  If no other planets are uncovered from the
first two OGLE campaigns, this implies a relative frequency of VHJ to
HJ of $r_{12}=0.15_{-0.07}^{+0.10}$, with a peak probability of $\sim
54\%$. In other words, RV and TR surveys are consistent
at better than the $1\sigma$ level.

\acknowledgments We would like to thank Dave Charbonneau, Debra
Fischer, Andrew Gould, Frederic Pont, Dimitar Sasselov, and Guillermo Torres for useful
discussions and comments.  We would also like the thank the anonymous
referee for a prompt report.  This work was supported by a Menzel Fellowship from the
Harvard College Observatory, by a Clay Fellowship from the Smithsonian
Astrophysical Observatory, by NSF-AST-0206278 and by the Carnegie
Institution of Washington.

\appendix

\section{Properties of the OGLE Campaigns}\label{ogleprops}

In this paper, we have focused on the OGLE TR surveys, and we briefly
summarize their properties here. OGLE mounted two separate campaigns
toward the Galactic bulge and disk.  In 2001, OGLE monitored 3 fields
toward the Galactic bulge over a period of $45\days$, with $793$
epochs per field taken on $\sim 32$ nights.  Approximately 52,000 disk
stars with RMS $<1.5\%$ light curves were searched for low-amplitude transits, yielding a total of $64$
candidates \citep{udalski02a,udalski02b}.  Of these candidates, one
planetary companion was confirmed with radial velocity follow-up
(OGLE-TR-56, \citealt{konacki03a}), and an additional two are planet
candidates with significant spectroscopic follow-up 
(OGLE-TR-10 and OGLE-TR-58, \citealt{konacki03b}).  In
2002, OGLE monitored an additional three fields in the Carina region of the
Galactic disk over a period of $95~\days$, with $\sim 1166$ epochs per
field taken on $\sim 76$ nights.  Approximately 103,000 stars 
with RMS $<1.5\%$ light curves were
searched for low-amplitude transits, yielding a total of $73$ candidates
\citep{udalski02c,udalski03}.  Of these candidates, three planetary
companions have been confirmed with radial velocity follow-up
(OGLE-TR-111, OGLE-TR-113, and OGLE-TR-132, \citealt{bouchy04,
konacki04, pont04}).

For both the 2001 and 2002 campaigns, candidates were found using the
BLS algorithm of \citep{kovacs02}.  This method works by folding
light curves about a trial period, and efficiently searching for dips
in the folded curves that have a $\sn$ larger than a given
threshold. \citet{udalski02b,udalski02c,udalski03} adopted a threshold
of $\sn=9$.  Figure \ref{fig:two} shows a contour of $\sn=9$ in the
$(R_p,P)$ plane, assuming $N_{\rm tot}=1166$ (as appropriate to the
2002 campaign), and $\sigma=0.005$, $R_*=\rsun$, and
$M_*=\msun$, as is typical of the OGLE target stars.

\begin{figure}
\epsscale{1.0} 
\plotone{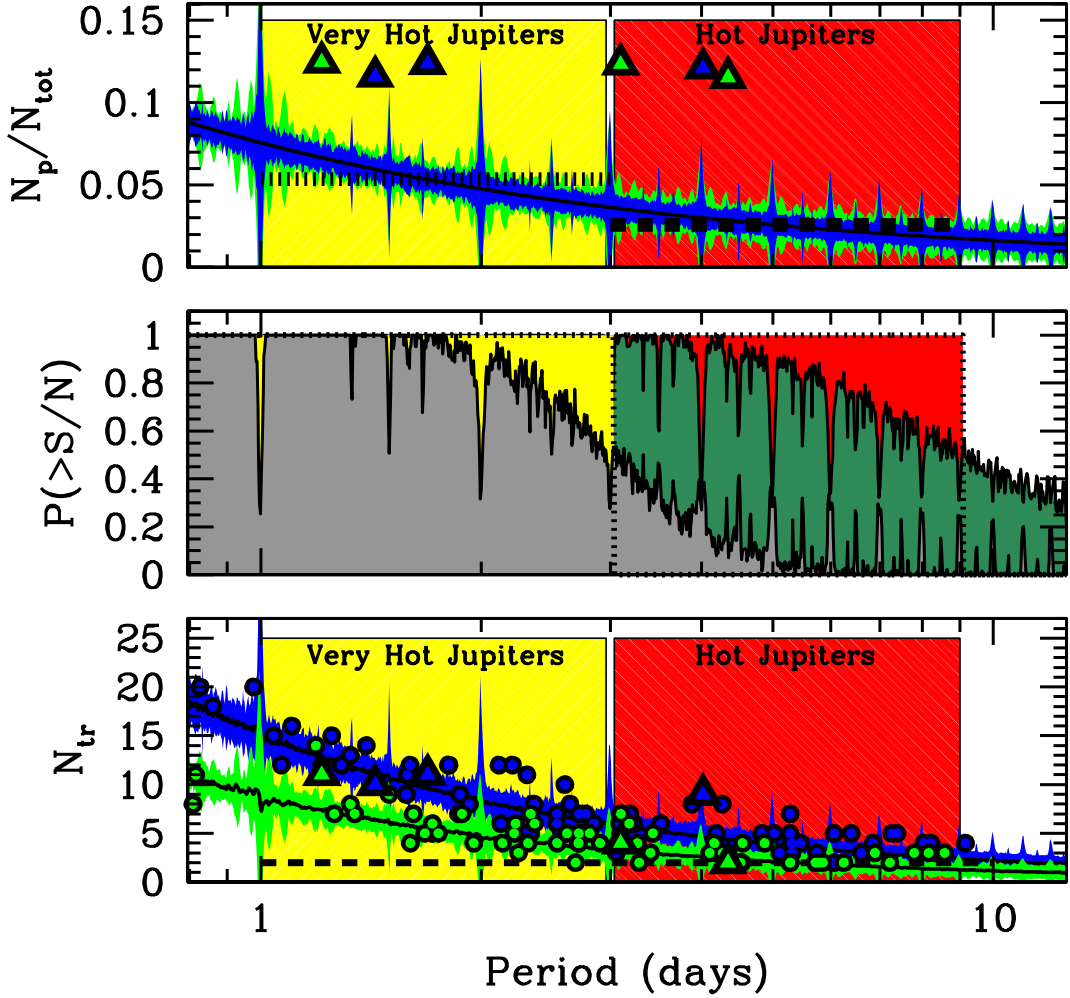}
\caption{\label{fig:four} 
Top Panel: The solid black line shows the mean fraction of data points
in transit, $N_p/N_{\rm tot}$, as a function of period $P$ for the OGLE transit (TR)
surveys, averaged over all phases.  The green and blue shaded regions
show the dispersion around the mean for the 2001 and 2002 OGLE
campaigns, respectively.  The yellow and red bands indicate the period
and mass ranges for our division into 
very hot Jupiters and hot Jupiters, respectively. The dotted and
dashed horizontal lines show the fraction of data points 
averaged over these two bins, assuming a uniform logarithmic
distribution in $P$.  The points show the periods of the OGLE
TR confirmed and candidate
planets; the ordinate values are arbitrary.  Green and blue points are
those found in the 2001 and 2002 campaigns, respectively.  
Middle Panel:  The green shaded curve shows the probability,
averaged over phase, for a planet with radius $R_p=0.1R_*$ 
to have $\sn>9$ for a photometric precision $\sigma_1=0.005$, for 
the 2002 OGLE campaign.  
The red box shows the naive expectation under uniform sampling
that all planets with $P\la 9\days$ will have $\sn>9$ for this
$\sigma<\sigma_1$.  
The grey curve shows the same probability for $\sigma_1=3^{1/3}\sigma_2\sim 0.007$;
three times more stars have $\sigma<\sigma_2$ than $\sigma <\sigma_1$.  The yellow
box shows the naive expectation that all planets with $P\la 3\days$
will have $\sn>9$ for $\sigma<\sigma_2$.
Bottom
Panel: The black lines show the number of observed transits, $N_{tr}$, 
with more
than three points for the given $P$ averaged over all phases.  The
green and blue shaded regions show the dispersion around the mean for
the 2001 and 2002 OGLE campaign, respectively.  The points show the
number of observed transits as a function of period for all the OGLE
low-luminosity or planetary transit candidates.  Green and blue points
are those found in the 2001 and 2002 campaigns, respectively.  Circles
show all candidates, triangles are the confirmed planets
or strong candidates.  The horizontal bar shows $N_{tr}=2$, the
minimum number of transits needed to establish a period.
}
\end{figure}

\section{Uneven Sampling and Finite Campaign Duration}\label{sec:aliasing}

In evaluating the relative sensitivity of transit surveys, we made the
simplistic assumption that the number of data points during transit is
proportional to the transit duty cycle for a central
transit, $N_p = R_*/\pi a$.  This
assumes random sampling and short periods as compared to the transit
campaign.  Of course, the OGLE campaigns have sampling that is far
from random, and in addition have finite durations of 1-3 months.
This introduces two effects.  First, the true fraction of points in
transit for an ensemble of light curves may be biased with respect to
the naive estimate of $N_p/\ntot =R_*/\pi a$.  In addition,
$N_p/\ntot$ will depend strongly on phase, and thus an ensemble of
systems at fixed $P$ will have a large dispersion in $N_p/\ntot$.

We illustrate the effects of the non-uniform sampling and finite
duration of the OGLE campaigns by analyzing the actual time stream of
one light curve from each of the 2001 and 2002 campaigns, namely
OGLE-TR-56 and OGLE-TR-113.  We fold each of these light curves about a
range of trial periods.  For each $P$, we choose a random phase, and
determine $N_p/\ntot$ assuming a primary of $M=\msun$ and $R=\rsun$.
We repeat this for many different phases, and determine the mean and
dispersion of $N_p/\ntot$.  The result is shown in Figure
\ref{fig:four}.  The mean agrees quite well with the naive estimate
of $N_p/\ntot = R_*/\pi a$.  However, the dispersion is significant,
with $\sigma_{N_p}/N_p$ ranging from $\sim 20\%$ for $P\sim 1\days$ to
$\sim 70\%$ for $P\sim 10\days$.  Since 
$\sn \propto N_p^{1/2}$, this translates to a dispersion in
$\sn$ of $\sim 0.5(\sigma_{N_p}/N_p) \sim 10\%-35\%$.  This implies
that, for a small number of samples (as is the case here), the value
of $N_p$ as a function of $P$ can have large stochastic variations
about the naive analytic estimate.  Such variations are largest
for near-integer day periods, as can by seen in Figure \ref{fig:four}.

The dispersion in the number of points during transit 
$N_p$ due to aliasing implies that the there is no longer a sharp
cutoff in the distance out to which one can detect a planet of a given 
period.  This is illustrated in
the middle panel of Figure \ref{fig:four},
where we plot the probability (averaged over phase)
that a planet with a fractional depth
$\delta=0.01$ will yield a $\sn>9$ as a function of $P$ for the
2002 campaign, assuming a 
photometric precision of $\sigma=0.005$ (green shaded curve).  Naively,
the uniform sampling approximation would imply that 
for a $\sn=9$ threshold all
planets should be detectable out to a period of $P \sim 9\days$, and none
with greater periods.  In fact, due to the dispersion in $N_p$ for fixed
period caused by aliasing, the transition is more gradual, such that it
is possible to detect planets with $P > 9~\days$, and there are
sharp dips in the completeness near integer day periods.  Figure 
\ref{fig:four} also shows the results for $\sigma=3^{-1/3}0.005\sim 0.007$ (grey shaded curve). There should
be three times more stars with $\sigma=0.007$ than $\sigma=0.005$, and
the naive expectation is that all planets with periods $P \la 3\days$ should
be detectable.  Clearly uneven 
sampling will affect the estimates of the relative sensitivity
of TR surveys as a function of period.

We note that OGLE-TR-111, which has $\delta\simeq 0.014$, $\sigma\sim
0.005$, and $P\simeq 4~\days$, would easily have exceeded the $\sn>9$
cut even under the assumption of uniform sampling, which would predict
$\sn \sim 16$.  Therefore, we find that it may not be necessary to
invoke aliasing to explain the detection of this planet, as suggested
by \citet{pont04}.  However, it is difficult to be definitive, because
the `by-eye' final selection of OGLE candidates may effectively impose a
$\sn$ limit that is significantly greater than the limit of $\sn>9$
used for the initial candidate selection.  The fact that a larger
number of transits ($N_{tr}=9$) were detected for OGLE-TR-111 than
would be expected based on its period {\it is} likely a consequence of its
near-integer period.

We can make a rough estimate of
the possible error made in adopting the naive estimate in the present
case by determining the expected distribution in the total
number of points in transit $N_p$.  We consider
our two fiducial period bins, $P_1=1-3\days$ and $P_2=3-9\days$, with
planets distributed uniformly in $\log P$ within each bin.  We then
draw three planets from each bin, with a random phase and period for
each planet.  We evaluate $N_p$ for each, and then find the mean $\ave{N_p}$
of the three planets.  We repeat this for many different realizations.
The ratio of the average $N_p$ for the two bins should be, on average,
$\ave{N_p}_1/\ave{N_p}_2 \simeq (1\days/3\days)^{-2/3} \sim 2$. For
the 2001 campaign, we find a median and $95\%$ confidence interval of
$\ave{N_p}_1/\ave{N_p}_2=2.11_{-1.06}^{+3.06}$, whereas for the 2002
campaign, we find $\ave{N_p}_1/\ave{N_p}_2=2.10_{-0.89}^{+1.83}$.  A
significant fraction of the variance arises from the small number of
samples; if we assume there is no dispersion of the relation between
$N_p$ and $P$ (i.e.\ uniform sampling), we find
$\ave{N_p}_1/\ave{N_p}_2=2.08_{-0.52}^{+0.69}$.  If we assume the
exact periods for the four confirmed planets and two candidates,
rather than random periods, we find very similar results, with
$\ave{N_p}_1/\ave{N_p}_2=1.87_{-0.67}^{+1.42}$ for the 2001 campaign,
and $\ave{N_p}_1/\ave{N_p}_2=1.92_{-0.77}^{+1.09}$ for the 2002
campaign.  

By incorporating these distributions of $\ave{N_p}_1/\ave{N_p}_2$ into
the analysis presented in \S\ref{sec:rvvtr}, it is possible to
determine the effect of aliasing on the inferred relative
frequency of VHJ to HJ.  We find that aliasing
does not alter our conclusions substantially.

\section{Radial Velocity Follow-up Biases}\label{rvfollowup}

One important distinction of TR surveys from RV surveys is that
candidate transiting planets must be confirmed by RV measurements.
Additional selection effects can be introduced at this stage.  We
discuss two such effects here.

The first effect is related to the detectability of the RV variations.
The detectability depends on the flux of the source and the magnitude
of the RV signal.  At fixed transit depth, shorter-period planet
candidates are, on average, fainter than longer-period candidates,
since $F_{\rm min} \propto P^{2/3}$.  
For photon-limited measurements, the typical RV error is $\sigma_{\rm RV} \propto
F_{\rm min}^{-1/2} \propto P^{-1/3}$.  
Thus, shorter period planets will
therefore require longer integration times to achieve a fixed 
$\sigma_{\rm RV}$. However,
the RV signal varies as $K \propto M_p P^{-1/3}$, and therefore, for all
else equal, the dependence of the relative signal-to-noise
$\sn=K/\sigma_K$ on period cancels out.  Thus, for fixed observing
conditions, the relative $\sn$ of RV measurements for VHJ versus HJ
depends (on average) only on their masses.  As discussed in \S\ref{sec:imp}, it
appears that the upper mass threshold of VHJ and HJ are different:
whereas there exists a real paucity of HJ with mass $\ga \mj$, the
four known VHJs all have masses $\ga \mj$.  This favors the
confirmation of VHJs.

An additional bias arises because two or more transits are needed to
establish the period of the planet.  Since an accurate period is
generally required for follow-up\footnote{Indeed, \citet{konacki03b}
rejected all OGLE candidates with only one transit detection as
unsuitable for follow-up.} (because
prior knowledge of the planet phase is important for efficient targeted
RV observations), and longer periods are less likely to
exhibit multiple transits, this bias also favors the confirmation of
short-period planets.  Figure \ref{fig:four} shows the mean and
dispersion of the number of transits with more than three data points
per transit
as a function of period for the 2001 and 2002 OGLE campaign.  The
majority of planets with periods of $P\la 3\days$ will exhibit at
least two transits, whereas planets with $P\ga 3\days$ are
increasingly likely to exhibit only one transit (or no
transits at all).

In summary, biases involved in both detection and confirmation of
transiting planets generally favor short-period planets.  It is
important to stress that all of the above arguments are true only on
average.  For the handful of planets currently detected, stochastic
effects associated with the small sample size change the magnitude or
even sign of the biases.

\end{document}